\documentclass[final,3p,times]{elsarticle}

\usepackage{amssymb}
\usepackage[english]{babel}
\usepackage[utf8x]{inputenc}
\usepackage{amsmath}
\usepackage[overload]{empheq}
\usepackage{framed}
\usepackage{marginnote}

\allowdisplaybreaks

\biboptions{numbers,sort&compress}

\usepackage{todonotes}
\usepackage{multirow,booktabs}
\usepackage{graphicx}
\usepackage{amsfonts}
\graphicspath{{images/}}
\usepackage{subfigure}
\usepackage[justification=centering]{caption}
\usepackage{epstopdf}
\usepackage[flushleft]{threeparttable}
\usepackage{balance}
\usepackage{bm}

\usepackage{float}

\usepackage{algorithmic}
\usepackage{algorithm}

\allowdisplaybreaks

\newtheorem{lemma}{Lemma}

\newtheorem{assumption}{Assumption}
\newtheorem{theorem}{Theorem}
\newtheorem{remark}{Remark}

\journal{Systems \& Control Letters}

\begin{document}

\begin{frontmatter}





\title{\LARGE {\bf When a New Robust Leader-Follower Tracking Problem Meets  Observer-Based Control} }

\title{\LARGE {\bf
A New Encounter Between  Leader-Follower Tracking and Observer-Based Control: Towards Enhancing Robustness  against Disturbances
}
}

\author[label1]{Chuan~Yan}\author[label1]{Huazhen~Fang\corref{cor1}}
\address[label1]{Department of Mechanical Engineering, University of Kansas, Lawrence, KS, 66045 USA}


 \cortext[cor1]{Corresponding author. E-mail address: fang@ku.edu.}



\begin{abstract}
This paper studies  robust tracking control for a leader-follower multi-agent system (MAS) subject to  disturbances.  A challenging problem is considered here, which differs from those in the literature in two aspects. First, we consider  the case when all the leader and follower agents are affected by disturbances, while the existing studies  assume only the followers to suffer disturbances. Second, we assume the disturbances  to be bounded only in rates of change rather than magnitude as in the literature. To address this  new problem, we propose a novel observer-based distributed tracking control design. As  a distinguishing feature,  the followers can cooperatively estimate  the  disturbance affecting the leader  to adjust their maneuvers accordingly, which is enabled by the design of  the  first-of-its-kind distributed disturbance observers. We build specific tracking control approaches for both first- and second-order MASs and prove that they can lead to bounded-error tracking, despite the challenges  due to the relaxed assumptions about disturbances. We further perform simulation to validate the proposed approaches.


\end{abstract}


\end{frontmatter}



\section{Introduction}\label{Sec:Introduction}




A leader-follower multi-agent system (MAS) refers to an MAS in which a group of follower agents   perform distributed  control while interchanging information with their neighbors to collectively track the state of a leader agent. A large body of work has been developed recently to deal with the tracking control design under diverse challenging situations, e.g., complex dynamics, communication delays, noisy measurements,  switching topologies, and limited energy budget, see~\cite{wen:2014:IJSS,shi:2009:AUT,hong:2008:distributed,li:2010:ITCS,zhu:2010:AUT,ya:2015:CDC,
ma:2010:JSSC,nourian:2012:TAC,ji:2006:ACC,bondhus:2005:cdc,defoort:2015:iet,Wu:SCL:2011,dimarogonas:2009:SCL,Mou:AUTO:2015,Yan:CEP:2018,Yan:IJC:2019,Yan:ACC:2018,Yan:ACC:2018b} and the references therein. However, a problem that has received inadequate attention to date is the case when the agents are subjected to disturbances. In a real world, disturbances can result from unmodeled dynamics, change in ambient conditions, inherent variability of the dynamic process, and sensor noises.  They can cause degradation and even failure of tracking control if not well addressed.

 A lead is taken in~\cite{hong:2008:distributed} with the study of disturbance-robust leader-follower tracking. It presents a distributed control design  that achieves tracking with a bounded error when magnitude-bounded disturbances affect the followers. This notion is extended in~\cite{li:2011:AUT} to make the followers affected by disturbances enter a bounded region centered around the leader in finite time. Another  finite-time tracking control approach is offered in~\cite{Zhang:IJC:2013}, where the sliding mode control technique is used to suppress the effects of disturbances. It is noted that, while the control designs in these works yield  robustness, they are  based on  upper bounds of the disturbances.
By contrast, a different   way is to capture the disturbances  by designing some observers and then adjust the control run based on the disturbance estimation. Obtaining an explicit knowledge of disturbances, this approach can advantageously reduce conservatism  in control and thus enhance the tracking performance further.
In~\cite{cao:2015:SCL,Sun:TCNS:2016}, disturbance observers are developed and integrated into tracking controllers such that a follower can estimate and offset the local disturbance interfering with its dynamics during tracking. The results in both studies point to the effectiveness of disturbance observers for improving  tracking accuracy --- for instance, the tracking errors can approach zero despite non-zero disturbances under certain   conditions.
 However, other than these two, there are no more  studies  on this subject  to the best of our knowledge. This leaves many  problems still open. Meanwhile, the potential of the disturbance-observer-based  approach    is still far from being fully explored. It is noteworthy that observer-based tracking control  has been investigated in a few works, e.g.,~\cite{hong:2008:distributed,li:2011:AUT,zhao:2013:SCL,zhao:2015:SCL,Du:IJC:2012}, but  observers in these studies are meant to infer various state or input variables rather than disturbances.

In this study, we uniquely focus on an open problem: can we enable distributed tracking control when not only the followers but also the leader are affected by unknown disturbances and when only the rates of change   of the disturbances are bounded?  The state of the art, e.g.,~\cite{hong:2008:distributed,li:2011:AUT,Zhang:IJC:2013,cao:2015:SCL,Sun:TCNS:2016}, generally considers that  disturbances   plague just the followers and that they are bounded in magnitude or approach fixed values as time goes by. The leader's dynamics, however, can also involve disturbances from a practical viewpoint. For example, consider an MAS composed of a few mobile ground robots, the changes in the slope of the road act as disturbances on every robot including the leader. The same can be said for the  wind affecting a group of unmanned aerial vehicles.  Such disturbances are more difficult to be rejected because the leader cannot measure them and share the information with any of the followers. Therefore, the tracking performance may be damaged when this occurs. Furthermore, it is usually desirable to relax the assumptions about disturbances to enhance the practical robustness of the control design. In~\cite{li:2011:AUT,Zhang:IJC:2013,cao:2015:SCL,Sun:TCNS:2016} , the disturbances are assumed to be bounded in magnitude. However, we wish to require the disturbances to be bounded in rates of change. This relaxation will be realistically  beneficial  for dealing with large disturbances but  also present more complexity to capture and suppress  the disturbances. It must be pointed out that the observer designs in~\cite{hong:2008:distributed,li:2011:AUT,Zhang:IJC:2013,cao:2015:SCL, Sun:TCNS:2016,zhao:2013:SCL,zhao:2015:SCL,Du:IJC:2012} cannot be extended to deal with the considered problem, due to the more challenging presence and nature of the disturbances. Hence, a solution is still absent from the literature.


To address the above problem, we  develop a novel observer-based distributed tracking control framework, which is the main contribution of this paper. Different from the previous studies, this framework builds on the notion that  a follower can gain a real-time awareness of not only its own but also the leader's dynamics through distributed estimation. We hence design a set of new observers and particularly, distributed disturbance observers that, for the first time, can enable  the  followers to collectively infer  the disturbance affecting the leader.  We perform the  observer-based tracking control  design for both first- and second-order MASs.
We then conduct theoretical analysis of the proposed approaches. We show that, even though disturbances are imposed on all the agents,   the tracking errors are still upper bounded (bounded-error tracking) as long as  the rates of change  of the disturbances  are bounded. Further, the tracking errors will approach zero (zero-error tracking) if the  disturbances converge to certain fixed points. We finally present  simulations to
validate the proposed approaches.

The rest of this paper is organized as follows. Section~\ref{Notation-Preliminaries} introduces  some preliminaries. Section~\ref{first-order} considers a leader-follower MAS with first-order dynamics, develops an observer-based distributed tracking control approach, and analyzes its performance rigorously. Section~\ref{second-order} proceeds to study the   second-order MASs and develops   a more sophisticated tracking control approach. Numerical simulation is offered in Section~\ref{simulation} to illustrate the effectiveness of the proposed results. Finally, Section~\ref{conclusion} gathers our
concluding remarks.

\section{Preliminaries}\label{Notation-Preliminaries}

This section introduces notation and basic concepts about graph theory and nonsmooth analysis.

\subsection{Notation}
The notation used throughout this paper is standard. The $n$-dimensional Euclidean space is denoted as $\mathbb{R}^n$. For a vector, $\|\cdot\|_1$ denotes the 1-norm, and $\|\cdot\|$ stands for the 2-norm. The notation $\mathbf 1$ represents a column vector of ones.
We let $\mathrm{diag}(\ldots)$ and $\det(\cdot)$  represent a block-diagonal matrix and the determinant of a matrix, respectively. The eigenvalues of an $N\times N$ matrix are $\lambda_i(\cdot)$ for $i=1,2,\ldots,N$. The minimum and maximum eigenvalues of a real, symmetric matrix are denoted as $\underline{\lambda} (\cdot)$ and $\bar{\lambda} (\cdot)$.  Matrices are assumed to be
compatible for algebraic operations if their dimensions are not explicitly stated. A $\mathcal{C}^k$ function is a function with $k$ continuous derivatives.

\subsection{Graph Theory}

We use a graph to describe the  information exchange topology for a leader-follower MAS. First, consider a network composed of $N$ independent followers, and model the interaction topology  as an undirected graph. The follower graph then is expressed as $\mathcal{G}=(\mathcal{V},\mathcal{E})$, where $\mathcal{V}=\{1, 2, \cdots, N\}$ is the vertex set and $\mathcal{E}\subseteq \mathcal{V}\times \mathcal{V}$ is the edge set  containing unordered pairs of vertices. A path is a sequence of connected edges in a graph. The follower graph is connected if there is a path between every pair of vertices.
The neighbor set of agent $i$ is denoted as $\mathcal{N}_i$, which includes all the agents in communication with it.
The adjacency
matrix of $\mathcal{G}$ is  $A=[a_{ij}]\in\mathbb{R}^{N \times N}$, which has non-negative elements. The element $a_{ij}>0 $  if  and only if $(i,j)\in
\mathcal{E}$, and moreover,
$a_{ii}=0$ for all $i \in \mathcal{V}$.  For the Laplacian matrix
$L=[l_{ij}]\in\mathbb{R}^{N \times N}$, $l_{ij}=-a_{ij}$ if $i\neq j$ and $ l_{ii}=\sum_{k\in \mathcal{N}_{i}}a_{ik}$.
The leader is numbered as vertex $0$ and can send information to its neighboring followers. Then, we have a graph $\bar{\mathcal{G}}$ , which consists of graph $\mathcal{G}$, vertex $0$ and edges from the leader   to its neighbors. The leader is globally reachable in $\bar{\mathcal{G}}$ if there is a path in graph $\bar{\mathcal{G}}$ from every vertex $i$ to vertex 0.
To express the graph $\bar{\mathcal{G}}$ more precisely, we denote the leader adjacency matrix associated with $\bar{\mathcal{G}}$ by $B=\mathrm{diag}(b_1,\ldots,b_N)$, where $b_i>0$ if the leader is a neighbor of agent $i$ and $b_i=0$ if it is not. The following lemmas will be useful.

\begin{lemma}\cite[Lemma 1.1]{ren:2010:SSBM}\label{L-notation}
The Laplacian matrix $L(\mathcal{G})$ has at least one zero eigenvalue, and all the nonzero eigenvalues are
positive. Furthermore, $L(\mathcal{G})$ has a simple zero eigenvalue and all the other eigenvalues are positive if and only if $\mathcal{G}$ is connected.
\end{lemma}

\begin{lemma}\cite[Lemma 4]{hu:2007:PASMA}\label{H-notation}
The matrix $H=L+B$ is positive stable
if and only if vertex 0 is globally reachable in $\mathcal{\bar{G}}$.
\end{lemma}

\subsection{Nonsmooth Analysis}
Consider the following discontinuous dynamical system
\begin{align}\label{Disconti-Dynamic}
\dot{x}=f(x), \ x\in \mathbb{R}^n,\ x(0)=x_0\in \mathbb{R}^n,
\end{align}
where $f(x): \mathbb{R}^n \rightarrow \mathbb{R}^n$ is defined almost everywhere (a.e.). In other words, it is defined   everywhere for $x\in \mathbb{R}^n\setminus W$, where $W$ is a subset of $\mathbb{R}^n$ of Lebesgue measure zero. Moreover, $f(x)$ is Lebesgue measurable in an open region and locally bounded. A vector variable $x(\cdot)\in \mathbb{R}^n$ is a Filippov solution of~\eqref{Disconti-Dynamic} on $[t_0,t_1]$ if $x(\cdot)$ is absolutely continuous on $[t_0,t_1]$ while for almost all $t\in [t_0,t_1]$, satisfying the following differential inclusion:
\begin{align}\label{Soluti}
\dot{x}\in \mathcal{K}[f](x)\triangleq \bigcap_{\delta>0}\bigcap_{\mu(M)=0}\mathrm{co}\{f(B(x,\delta)\setminus M)\},
\end{align}
where $\bigcap_{\mu(M)=0}$ represents the intersection over all sets $M$ of Lebesgue measure zero, $\mathrm{co}(\cdot)$ denotes the closure of a  convex hull and $B(x,\delta)$ denotes an open ball of radius $\delta$ centered at $x$.
Let $V(x):\mathbb{R}^n \rightarrow \mathbb{R}$ be a locally Lipschitz continuous function. Its Clarke's generalized gradient is given by
\begin{align*}
\partial V(x)\triangleq \mathrm{co}\left\{\lim_{i\rightarrow \infty} \nabla V(x_i)|x_i \rightarrow x, x_i \notin \Omega_V \cup M\right\},
\end{align*}
where $\nabla V(x)$ is the conventional gradient, and $\Omega_V$ denotes a set of Lebesgue measure zero which includes all points where $\nabla V(x)$ does not exist. Moreover, the set value of the derivative of $V$ associated with~\eqref{Disconti-Dynamic} is defined as
\begin{align*}
\mathcal{L} \dot{V}(x)=\{a\in \mathbb{R}\}| \exists \ v\in \mathcal{K}[f](x)\ \mbox{such that} \ \zeta\cdot v=a, \forall\zeta\in
\partial V(x)\}.
\end{align*}
The following lemma  will be used later.

\begin{lemma}\cite[Page 32]{clarke:1983:NY}\label{Filippov}
Let $x(t):[t_0,t_1]\rightarrow \mathbb{R}^n$ be a Filippov solution of~\eqref{Soluti}. Let $V(x)$ be a locally Lipschitz and regular function. Then $d/dt(V(x(t)))$ exists a.e. and $d V(x(t)) /dt\in \mathcal{L} \dot{V}(x)$ a.e.
\end{lemma}

\subsection{Assumptions}
Throughout this paper, we consider  a leader-follower MAS with $N+1$ agents. The agents are numbered sequentially. The one numbered as 0 serves as  the leader, and the other agents are followers. Each agent is driven by an input $u_i$ and simultaneously affected by an external disturbance $f_i$ for $i=0,1,\ldots,N$. We make the following  assumptions.
\begin{assumption}\label{Asmp_Leader_Input}
The input $u_0\in \mathcal{C}^1$ has a bounded first-order derivative, satisfying $|\dot{u}_0|\leq w$, where $w$ is unknown.
\end{assumption}

\begin{assumption}\label{Asmp_Disturbance}
The external disturbance $f_i$ for $i=0,1,\ldots,N$ has a bounded first-order derivative, i.e., $\| \dot f_0 {\mathbf 1}_{N\times 1} \| \leq q_0 $ and $\left\|  \begin{bmatrix} \dot f_1 & \dot f_2 & \cdots & \dot f_N  \end{bmatrix}^\top \right\| \leq q_1 $, where $ q_0,q_1 \geq 0$.
\end{assumption}

\section{First-Order Leader-Follower Tracking}\label{first-order}

This section studies first-order leader-follower tracking  with   disturbances. We develop an observer-based control approach, pivoting the design on a set of   observers to make a follower aware of   the leader's and its own disturbances. We further analyze the closed-loop stability of the proposed approach.

\subsection{Problem Formulation}

Consider an MAS with $N+1$ agents, in which agent  0 is the leader and the others are  followers. An agent's dynamics is given by
\begin{align}\label{follower-dynamics-first}
\dot{x}_{i}=u_{i}+f_{i},\quad x_{i}, u_i, f_i \in \mathbb{R}, \quad i=0,1,\ldots,N,
\end{align}
where $x_{i}$ is the position, $u_{i}$ the control input equivalent to the velocity maneuver, and $f_{i}$ the unknown disturbance. Suppose that Assumptions~\ref{Asmp_Leader_Input}-\ref{Asmp_Disturbance} hold. Here, the objective is to design a distributed control law for $u_i$ such that each follower can control its dynamics to track the leader's trajectory   {\em via} exchanging information with its neighbors.

\begin{remark}
Compared with previous studies, the problem setting here is more generic and applicable to a wide range of practical scenarios. Below, we  outline a  comparison with~\citep{hong:2008:distributed,li:2011:AUT,Zhang:IJC:2013,cao:2015:SCL,Sun:TCNS:2016}, which are the main references about tracking control with disturbances and henceforth  referred to as the existing literature.
First, this work considers an input-driven leader, while   the leader is usually assumed to be input-free in the literature.  Assumption~\ref{Asmp_Leader_Input} only requires  the leader's input to be bounded in rate of change (with the bound unknown), which can  be easily satisfied since practical actuators  only allow limited ramp-ups.
Second, Assumption~\ref{Asmp_Disturbance} imposes disturbances   on all the leader and follower agents,  while the literature assumes only followers to be affected by disturbances. Note that the case when a disturbance is inflicted on the leader is nontrivial. This is because the leader's disturbance is very difficult to be determined by the followers, especially in a distributed network where many followers cannot directly interchange information with the leader.
Further,  the disturbances are assumed to have only bounded rates of change rather than bounded magnitude as required in the literature. This can be greatly useful for dealing with  very large disturbances. From the comparison, we conclude that the considered problem     is less restrictive than the predecessors, which still remains an open challenge.  \hfill$\bullet$
\end{remark}

\subsection{Proposed Algorithm}

Given  the above problem setting, we propose an observer-based tracking control approach. The development begins with the design of a distributed linear continuous controller  for a follower (say, follower $i$).  It   crucially incorporates the estimation of three unknown variables, $u_0$, $f_0$ and $f_i$,     enabling follower $i$ to maneuver through simultaneously emulating the  input  and disturbance driving the leader  and    offsetting  the local disturbance.
We subsequently construct three observers     to achieve the estimation to be integrated  with the controller.  

Considering follower $i$, we propose to design its controller as follows:
\begin{align}\label{controller-first}
u_i=&-k \left[\sum_{j \in \mathcal{N}_{i}}a_{ij}(x_i-x_j)
+b_{i}(x_i-x_0)\right]+\hat{u}_{0,i}+\hat{f}_{0,i}-\hat{f}_i,
\end{align}
where $k >0$ is the control gain, $\hat{f}_{0,i}$ and $\hat{u}_{0,i}$ are follower $i$'s respective estimates of the leader's disturbance $f_0$ and input $u_0$, and $\hat{f}_i$ is follower $i$'s estimate of its own disturbance $f_i$. Note that $b_i>0$ if the leader is agent $i$'s neighbor and $b_i=0$ if it is not.
In~\eqref{controller-first}, the term $-\sum_{j \in \mathcal{N}_{i}}a_{ij}(x_i-x_j)
-b_{i}(x_i-x_0)$ is employed to drive  follower $i$ approaching the leader; the term $\hat{u}_{0,i}+\hat{f}_{0,i}$ ensures that  follower $i$ applies maneuvers consistent with the leader's input and disturbance; the term $-\hat{f}_i$ is used to cancel the local disturbance. For  this controller, we build a series of observers as shown below to estimate $u_0$, $f_0$ and $f_i$, respectively.

To begin with we propose an observer as follows to obtain ${\hat{u}}_{0,i}$, which is based on  the design   in~\cite{Yan:IJC:2019}:
\begin{subequations}\label{u0-estimation-first}
\begin{align}
\dot{\hat{u}}_{0,i}&=-\sum_{j \in \mathcal{N}_{i}}a_{ij}(\hat{u}_{0,i}-\hat{u}_{0,j})-b_{i}
(\hat{u}_{0,i}-u_{0})
-d_i\cdot\mathrm{sgn}\left[\sum_{j \in \mathcal{N}_{i}}a_{ij} \label{u0-estimation-first-a}
(\hat{u}_{0,i}-\hat{u}_{0,j})+b_{i}
(\hat{u}_{0,i}-u_{0})\right],\\  \label{adaptive-gain}
\dot{d}_i&=\tau_{i}\left|\sum_{j \in   \mathcal{N}_{i}}a_{ij}(\hat{u}_{0,i}-\hat{u}_{0,j})+b_{i}(\hat{u}_{0,i}-u_{0})\right| ,
\end{align}
\end{subequations}
for $i=1,2,\ldots,N$, where $d_i$ is the observer gain and $\tau_i>0$ a scalar coefficient. For~\eqref{u0-estimation-first-a}, the leading term on the right-hand side, $-\sum_{j \in \mathcal{N}_{i}}a_{ij}(\hat{u}_{0,i}-\hat{u}_{0,j})
-b_{i}(\hat{u}_{0,i}-u_{0})$, is used to make $\hat{u}_{0,i}$ approach $u_0$; the signum function $\mathrm{sgn}(\cdot)$ is aimed to overcome the effects of $u_0$'s first-order dynamics, i.e., $\dot{u}_0$, and ensure the convergence of $\hat u_{0,i}$ to $u_0$. Note that the observer gain, $d_i$, is adaptively adjusted through~\eqref{adaptive-gain}. As such, a reasonable gain can be determined  even if the upper bound of $\dot u_0$, $w$, is unknown (see Assumption~\ref{Asmp_Leader_Input}).

The following disturbance observer is proposed for follower $i$ to estimate $f_0$:
\begin{subequations}\label{f0-estimation-first}
\begin{align}
&\dot{z}_{f0,i}=-b_{i}z_{i}-b_{i}^{2}x_{0}-\sum_{j \in \mathcal{N}_{i}}a_{ij}(\hat{f}_{0,i}-\hat{f}_{0,j})-b_iu_0 , \\
&\hat{f}_{0,i}=z_{f0,i}+b_{i}x_{0},
\end{align}
\end{subequations}
where $z_{f0,i}$ is the internal state. The design of~\eqref{f0-estimation-first} is inspired by~\cite{yang:2013:ITIE}, in which   a centralized  disturbance observer    is developed for a single plant. Here,  transforming the original design, we  build the distributed observer as above such that   follower $i$ can  estimate $f_0$ in a distributed manner. 

The last observer, designed as follows, enables follower $i$ able to infer the disturbance $f_i$ inherent in its own dynamics:
\begin{subequations}\label{fi-estimation-first}
\begin{align}
&\dot{z}_{f,i}=-lz_{f,i}-l^{2}x_{i}+u_i, \\
&\hat{f}_{i}=z_{f,i}+lx_{i}.
\end{align}
\end{subequations}
Here, $l>0$ is the observer gain, and $z_{f,i}$ is this observer's internal state. 

Combining~\eqref{controller-first}-\eqref{fi-estimation-first}, we obtain a complete description of an observer-based distributed tracking controller. Next, we will analyze its  closed-loop stability.

\subsection{Stability Analysis}

Define $e_{u,i}=\hat{u}_{0,i}-u_0$, which is the input estimation error. According to~\eqref{u0-estimation-first}, the closed-loop dynamics of $e_{u,i}$ can be written as
\begin{align*}
\dot{e}_{u,i}
&=-b_ie_{u,i}-\sum_{j \in \mathcal{N}_{i}}a_{ij}(e_{u,i}-e_{u,j})
-d_i\cdot\mathrm{sgn}\left[\sum_{j \in \mathcal{N}_{i}}a_{ij}(e_{u,i}-e_{u,j})+b_ie_{u,i}\right]-\dot{u}_0.
\end{align*}
Further, let us concatenate $e_{u,i}$ for $i=1,2,\ldots,N$ and define $e_u=\left[\begin{matrix} e_{u,1}& e_{u,2}&\cdots& e_{u,N}\end{matrix}\right]^{\top}$. The dynamics of $e_u$ can be expressed as
\begin{align}\label{eu-dynamics-first}
\dot{e}_{u}=-He_u-D\cdot\mathrm{sgn}(He_u)-\dot{u}_0\mathbf{1},
\end{align}
where $H=B+L$ and $D=\mathrm{diag}(d_{1},d_{2},\ldots, d_{N})$. It is noted that the  signum-function-based term at the right-hand side of~\eqref{eu-dynamics-first} is discontinuous, measurable and locally bounded. Hence, there exists a Filippov solution to~\eqref{eu-dynamics-first}, which is represented by a differential inclusion as follows:
\begin{align*}\label{eu-dynam}
\dot{e}_{u}\in \mathcal{K}\left[-He_u-D\cdot\mathrm{sgn}(He_u)-\dot{u}_0\mathbf{1}\right].
\end{align*}
The following lemma characterizes  the convergence of $e_u$.

\begin{lemma}\label{u0-analysis}
If Assumption~\ref{Asmp_Leader_Input} holds, then
\begin{align}\label{u0_estimation_convergence}
\lim_{t \rightarrow \infty} |e_{u,i}| =0,
\end{align}
for $i=1,2,\ldots,N$.
\end{lemma}

{\noindent\bf Proof:}
By Lemmas~\ref{L-notation} and~\ref{H-notation}, $H$ is positive definite. For~\eqref{eu-dynamics-first}, consider
\begin{align*}
\bar{V}_1(e_u)=\frac{1}{2}e_u^\top He_u,\
\tilde{V}_1=\sum_{i=1}^N \frac{(d_i-\beta)^{2}}{2\tau_i},
\end{align*}
where   $\beta\geq w$, as noted. We then take $V_1=\bar{V}_1(e_u)+\tilde{V}_1$ as a Lyapunov functional candidate.
For the set-valued Lie derivative of $\bar{V}_1(e_u)$, we have
\begin{align*}
\mathcal{L}\dot{\bar{V}}_1&=\mathcal{K}\left[-e_u^\top H^2e_u-e_u^\top H D\cdot \mathrm{sgn}(He_u)-e_u^\top H\dot{u}_0\mathbf{1}\right]\\ \nonumber
&= \mathcal{K}\Big[-\sum_{i=1}^N d_i\left(\sum_{j \in \mathcal{N}_{i}}a_{ij}(\hat{u}_{0,i}-\hat{u}_{0,j})+b_{i}(\hat{u}_{0,i}-u_{0})\right)^\top\\ \nonumber
&\quad \cdot\mathrm{sgn}\left(\sum_{j \in \mathcal{N}_{i}}a_{ij}(\hat{u}_{0,i}-\hat{u}_{0,j})+b_{i}(\hat{u}_{0,i}-u_{0})\right)-e_u^\top H^2e_u-e_u^\top H\dot{u}_0\mathbf{1}\Big]\\ \nonumber
&\leq-\sum_{i=1}^N d_i\left|\sum_{j \in \mathcal{N}_{i}}a_{ij}(\hat{u}_{0,i}-\hat{u}_{0,j})+b_{i}(\hat{u}_{0,i}-u_{0})\right|
-e_u^\top H^2e_u+w\|He_u\|_1
\end{align*}
by the fact that $\mathcal{K}[f]=\{f\}$ if $f$ is continuous. Invoking Lemma~\ref{Filippov}, we obtain that $\dot{\bar{V}}_1\in \mathcal{L}\dot{\bar{V}}_1$. Then, the derivative of $V_1$ is given by
\begin{align*}
\dot{V}_1&=\dot{\bar{V}}_1+\dot{\tilde{V}}_1=\dot{\bar{V}}_1+\sum_{i=1}^N \frac{(d_i-\beta)\dot{d_i}}{\tau_i}\\ \nonumber
&\leq-\sum_{i=1}^N d_i\left|\sum_{j \in \mathcal{N}_{i}}a_{ij}(\hat{u}_{0,i}-\hat{u}_{0,j})+b_{i}(\hat{u}_{0,i}-u_{0})\right|
-e_u^\top H^2e_u+w\|He_u\|_1 \\ \nonumber
&\quad +\sum_{i=1}^N (d_i-\beta)\left|\sum_{j \in \mathcal{N}_{i}}a_{ij}(\hat{u}_{0,i}-\hat{u}_{0,j})+b_{i}(\hat{u}_{0,i}-u_{0})\right| \\
&=-e_u^\top H^2e_u-(\beta-w)\|He_u\|_1.
\end{align*}
It is noted that $e_u^\top H^2e_u\geq 0$. This, in addition to the fact that  there always exists a $\beta$ such that $\beta \geq w$ by Assumption~\ref{Asmp_Leader_Input}, ensures $\dot{V_1}\leq 0$. As a result, $V_1(e_u)$ is nonincreasing, which implies that
$e_u$ and $d_i$ are bounded. From~\eqref{adaptive-gain}, it follows that $d_i$ is monotonically increasing,   indicating that   $d_i$ should converge  to some finite value. In the meantime, since  $V_1$ is nonincreasing and lower-bounded  by zero,  it  should approach a finite limit. Defining $s(t) =\int_0^t e_u^\top (\tau)H^2e_u(\tau) d \tau$, we have $s(t) \leq V_1(0)-V_1(t) $ by integrating $\dot{V}_1(e_u)\leq -e_u^\top H^2e_u$. Hence, $s(t)$ will also approach a finite limit.  Due to the boundedness of $e_u$ and $\dot e_u$, $\ddot s$ is also bounded. This implies that $\dot s$ is uniformly continuous. By Barbalat's Lemma~\citep{khalil:1996:PHNJ}, $\dot s(t) \rightarrow 0$ as $t\rightarrow \infty$, indicating that $e_u \rightarrow 0$ as $t \rightarrow \infty$. \hfill$\bullet$

Now, consider the distributed observer for  $f_0$. Define $e_{0f,i}=\hat{f}_{0,i}-f_0$, which is follower $i$'s estimation error for $f_0$. Using~\eqref{f0-estimation-first}, the  dynamics of $e_{0f,i}$ is given by
\begin{align*}
\dot{e}_{0f,i}
&=-b_ie_{0f,i}-\sum_{j \in \mathcal{N}_{i}}a_{ij}(\hat{f}_{0,i}-\hat{f}_{0,j})-\dot{f}_0.
\end{align*}
Then, defining $e_{0f}=\left[\begin{matrix} e_{0f,1}& e_{0f,2}&\cdots& e_{0f,N}\end{matrix}\right]^{\top}$, we have
\begin{align}\label{ef0-dynamics-first}
\dot{e}_{0f}=-He_{0f}-\dot{f}_0\mathbf{1}.
\end{align}
The following lemma reveals the upper boundedness of  $e_{0f}$ under  Assumption~\ref{Asmp_Disturbance}.

\begin{lemma}\label{f0-analysis}
If Assumption~\ref{Asmp_Disturbance} holds, then
\begin{subequations}
\begin{align}\label{e0f-boundedness}
\|e_{0f}(t)\| & \leq \|e_{0f}(0)\|+\frac{q_0}{\underline{\lambda}(H)}, \ \ t>0,\\ \label{e0f-limit}
\lim\limits_{t\rightarrow \infty} \|e_{0f}(t)\| & \leq \frac{q_0}{\underline{\lambda}(H)}.
\end{align}
\end{subequations}
\end{lemma}

{\noindent\bf Proof:}
Consider the Lyapunov function candidate $V_2(e_{0f})=\frac{1}{2}e_{0f}^\top e_{0f}$ for~\eqref{ef0-dynamics-first}. According to Assumption~\ref{Asmp_Disturbance}, we have
\begin{align*}
\dot{V_2}(e_{0f})&=-e_{0f}^\top He_{0f}-e_{0f}^\top \dot{f}_0\mathbf{1}
\leq -\underline{\lambda}(H)\|e_{0f}\|^{2}+\|e_{0f}\|\|\dot{f}_0\mathbf{1}\|
\leq -\underline{\lambda}(H)\|e_{0f}\|^{2}+q_0\|e_{0f}\|.
\end{align*}
The above inequality can be rewritten as
\begin{align*}
\dot{V_2} \leq -2\underline{\lambda}(H)V_2 +\sqrt{2}q_0\sqrt{V_2}.
\end{align*}
It then follows that
\begin{align}\label{V0f-infinity}
\sqrt{V_2 (t)}&\leq \sqrt{V_2 (0)}e^{-\underline{\lambda}(H)t}+\frac{\sqrt{2}q_0}{2\underline{\lambda}(H)}
\left(1-e^{-\underline{\lambda}(H)t}\right)\leq \sqrt{V_2(0)}+\frac{\sqrt{2}q_0}{2\underline{\lambda}(H)}.
\end{align}
Then,~\eqref{e0f-boundedness} can result from~\eqref{V0f-infinity} because $\sqrt{V_2}={\sqrt{2}\over 2} \|e_{0f}\|$. Meanwhile, for the first inequality in~\eqref{V0f-infinity}, taking the limits of both sides as $t\rightarrow \infty $ would yield~\eqref{e0f-limit}.
\hfill$\bullet$


For the estimation of $f_i$, define the  error as $e_{f,i}=\hat{f}_{i}-f_{i}$ and further the   vector $e_f=\left[\begin{matrix} e_{f,1} & e_{f,2}& \ldots &e_{f,N}\end{matrix}\right]^{\top}$. By~\eqref{fi-estimation-first}, the dynamics of $e_f$ is governed by
\begin{align}\label{fi_estimation_error_dynamics}
\dot{e}_f=-le_f-\dot{f},
\end{align}
where $\dot{f}= \begin{bmatrix} \dot f_1 & \dot f_2 & \cdots & \dot f_N  \end{bmatrix}^\top$.
The next lemma shows that the error $e_f$ is bounded under Assumption~\ref{Asmp_Disturbance}. Its proof is similar to that of Lemma~\ref{f0-analysis} and thus omitted here.

\begin{lemma}\label{f-analysis}
If Assumption~\ref{Asmp_Disturbance} holds,  then
\begin{align*}\label{ef_boundedness}
\|e_{f}(t)\| &\leq \|e_{f}(0)\|+\frac{q_1}{l}, \  \ t>0\\ 
\lim\limits_{t\rightarrow \infty} \|e_{f}(t)\| & \leq \frac{q_1}{l}.
\end{align*}
\end{lemma}

With the above results , we are now in a good position to characterize the properties of the tracking error.
Define follower $i$'s  tracking error as $e_i=x_i-x_0$, and put together $e_i$ for $i=1,2,\ldots,N$ to form the vector $e=\left[\begin{matrix} e_{1}&e_{2}&\cdots&e_{N}\end{matrix}\right]^{\top}$. Using~\eqref{follower-dynamics-first} and~\eqref{controller-first}, it can be derived that the dynamics of $e$  can be described as
\begin{align}\label{e-dynamics-first}
\dot{e}=-k He+e_{0f}+e_u-e_f.
\end{align}
The following theorem provides a key technical result.

\begin{theorem}\label{ui-analysis-first}
Suppose that Assumptions~\ref{Asmp_Leader_Input} and~\ref{Asmp_Disturbance} hold. Then,
\begin{subequations}
\begin{align}\label{e_boundedness}
&\|e(t)\|\leq \|e(0)\|+\frac{\|e_{0f}(0)\|+\|e_{u}(0)\|+\|e_{f}(0)\|+\frac{q_0}{\underline{\lambda}(H)}+
\frac{q_1}{l}}{k \underline{\lambda}(H)}, \\ \label{e_limit}
&\lim\limits_{t\rightarrow \infty}\|e\| \leq \frac{\frac{q_0}{\underline{\lambda}(H)}+
\frac{q_1}{l}}{k \underline{\lambda}(H)}.
\end{align}
\end{subequations}
\end{theorem}

{\noindent\bf Proof:}
Take the Lyapunov function candidate $V_3(e)=\frac{1}{2}e^{\top}e$ for~\eqref{e-dynamics-first}. Consider its  derivative:
\begin{align*}
\dot{V}_3&=-k e^\top H e+e^\top e_{0f}+e^\top e_{u}-e^\top e_{f}\\
&\leq -k \underline{\lambda}(H)\|e\|^{2}+\|e\| \cdot \|e_{0f}\|+\|e\| \cdot \|e_u\|+\|e\| \cdot \|e_f\|,
\end{align*}
where $\underline{\lambda}(H)>0$.
Equivalently, we have
\begin{align*}
\dot{V}_3\leq -2k \underline{\lambda}(H)V_3+\sqrt{2}(\|e_{0f}\|+\|e_u\|+\|e_f\|)\sqrt{V_3}.
\end{align*}
Then,
\begin{align*}
\sqrt{V_3(t)}&\leq \sqrt{V_3(0)}e^{-k \underline{\lambda}(H)t}
+\frac{\sqrt{2}(\|e_{0f}(t)\|+\|e_u(t)\|+\|e_f(t)\|)}{2k \underline{\lambda}(H)}
(1-e^{-k \underline{\lambda}(H)t})\\
&\leq \sqrt{V_3(0)}+\frac{\sqrt{2}(\|e_{0f}(t)\|+\|e_u(t)\|+\|e_f(t)\|)}{2k \underline{\lambda}(H)},
\end{align*}
which, based on Lemmas~\ref{u0-analysis}-\ref{f-analysis}, indicates~\eqref{e_boundedness}-\eqref{e_limit}.
\hfill$\bullet$


Theorem~\ref{ui-analysis-first} shows that the proposed   observer-based  controller can make each follower  track the leader  with bounded position errors  despite the disturbances. Therefore, we can say that the influence of the disturbances is effectively suppressed and that tracking is achieved in a practically meaningful  manner.

\begin{remark}\label{Perfect-error-tracking-first-order}
For the proposed controller, the tracking performance will be further improved if the disturbances satisfy some stricter conditions. In particular, it is noteworthy that perfect or zero-error tracking can be attained if the disturbances see their rates of change gradually settle down to zero, i.e., $\dot{f}_i(t) \rightarrow 0$ as $t \rightarrow \infty$ for $i=0,1,\ldots,N$. The proof can be developed following similar lines as above and  is omitted here. \hfill$\bullet$
\end{remark}

\section{Second-Order Leader-Follower Tracking}\label{second-order}

This section considers   leader-follower tracking control for a second-order MAS. An agent's dynamics now involves position, velocity, acceleration and disturbance:
\begin{align}\label{leader-follower-dynamics-second}
\left\{
\begin{array}{l}
\dot{x}_{i}=v_{i},\quad x_{i}\in \mathbb{R},\\
\dot{v}_{i}=u_{i}+f_i, \quad v_{i}\in \mathbb{R},
\end{array}
\right.
\end{align}
for $i=0,1,\ldots,N$. Here, $x_{i}$ is the position, $v_{i}$ the velocity, $u_{i}$ the acceleration input, and $f_i$ the  disturbance. Still, agent 0 is the leader, and the others are followers numbered from 1 to $N$. We continue to apply Assumptions~\ref{Asmp_Leader_Input}-\ref{Asmp_Disturbance}  here and   set  the   objective of making the followers achieve bounded-error tracking of the leader in the presence of the disturbances.

For a general problem formulation, we further assume  that   no velocity sensor is deployed on the leader and followers. Hence, there are no velocity measurements throughout the tracking process. The absence of the velocity information, together with the unknown disturbances, makes the tracking control  problem more complex  than in the first-order case, thus  requiring a substantial sophistication of the   observer-based control approach in Section~\ref{first-order}. Here, we will custom build an observer-based tracking controller and   develop new velocity and disturbance observers.

Consider follower $i$. We propose the following distributed controller:
\begin{align}\label{controller-second}
u_i=&-k \left[\sum_{j \in \mathcal{N}_{i}}a_{ij}(x_i-x_j)
+b_{i}(x_i-x_0)\right]
-(\hat v_i-\hat v_{0,i})+\hat{u}_{0,i}+\hat{f}_{0,i}-\hat{f}_i,
\end{align}
where $k >0$ is the control gain. In addition, $\hat u_{0,i}$, $\hat{v}_{0,i}$, $\hat f_{0,i}$, $\hat v_i$ and $\hat f_i$ are, respectively, follower $i$'s estimates of $u_0$, $v_0$, $f_0$, $v_i$ and $f_i$.  The terms $-\sum_{j \in \mathcal{N}_{i}}a_{ij}(x_i-x_j)-b_{i}(x_i-x_0)$ and $-(\hat v_i-\hat v_{0,i})$ are used to enable the follower to track the leader in both position and velocity; the term $\hat{u}_{0,i}+ \hat{f}_{0,i}$ is used to make the follower steer itself with a maneuvering input close to the combined input and disturbance driving the leader; the term $-\hat{f}_i$ is meant to offset the local disturbance.

With the above controller structure, we need to construct observers that can obtain the needed estimates. First, it is noted that the input observer of $u_0$ in~\eqref{u0-estimation-first} can be applied here without any change, and  its convergence property as shown in  Lemma~\ref{u0-analysis}  also holds in this case. Then, we develop the following observer   such that follower $i$ can estimate the leader's unknown velocity:
\begin{subequations}\label{v0-estimation}
\begin{align}
\dot{z}_{v0,i}=&-b_{i}z_{v0,i}-b_{i}^{2}x_{0}-\sum_{j \in \mathcal{N}_{i}}a_{ij}(\hat{v}_{0,i}-\hat{v}_{0,j})
+\hat{f}_{0,i}+\hat{u}_{0,i}, \\
\hat{v}_{0,i}=&z_{v0,i}+b_{i}x_{0},
\end{align}
\end{subequations}
where $z_{v0,i}$ is the internal state of this observer. Follower $i$'s observer for the leader's disturbance is then proposed as
\begin{subequations}\label{f0-estimation-2nd-order}
\begin{align}
&\dot{z}_{f0,i}=-z_{f0,i}-\hat{v}_{0,i}-\hat{u}_{0,i}, \\
&\hat{f}_{0,i}=z_{f0,i}+\hat{v}_{0,i}.
\end{align}
\end{subequations}
The next observer enables follower $i$ to estimate its own velocity:
\begin{subequations}\label{vi-estimation}
\begin{align}
&\dot{z}_{v,i}=-lz_{v,i}-l^{2}x_{i}+\hat{f}_{i}+u_i, \\
&\hat{v}_{i}=z_{v,i}+lx_{i}.
\end{align}
\end{subequations}
Here, $l>0$ is the gain for this observer, and $z_{v,i}$  the internal state. The final observer is aimed to allow follower $i$ to infer its local disturbance. It is designed as
\begin{subequations}\label{fi-estimation-2nd-order}
\begin{align}
&\dot{z}_{f,i}=-z_{f,i}-\hat{v}_{i}-u_i, \\
&\hat{f}_{i}=z_{f,i}+\hat{v}_{i},
\end{align}
\end{subequations}
where $z_{f,i}$ is the internal state.

From above, a complete observer-based distributed tracking controller can be built by putting together the control law~\eqref{controller-second} and the observers in~\eqref{u0-estimation-first} and~\eqref{v0-estimation}-\eqref{fi-estimation-2nd-order}.
The following theorem is the main   result about the closed-loop stability of the proposed controller.
\begin{theorem}\label{ui-analysis-second}
Suppose that Assumptions~\ref{Asmp_Leader_Input}-\ref{Asmp_Disturbance} hold and apply the proposed distributed tracking controller given in~\eqref{u0-estimation-first} and~\eqref{controller-second}-\eqref{fi-estimation-2nd-order} to the MAS in~\eqref{leader-follower-dynamics-second}. 
Then, there exist    $\delta>0$ and $\epsilon >0 $ such that
\begin{subequations}
\begin{align}\label{position-velocity-tracking-boundedness}
 \left\| \left[\begin{matrix} x_i(t)-x_0(t)\\v_i(t)-v_0(t)   \end{matrix}\right] \right\| &\leq \delta, \ \ t>0, \\ \label{position-velocity-tracking-bounded-limit}
\lim\limits_{t\rightarrow \infty} \left\| \left[\begin{matrix} x_i(t)-x_0(t)\\v_i(t)-v_0(t)   \end{matrix}\right] \right\| & \leq \epsilon ,
\end{align}
\end{subequations}
for $i = 1,2,\ldots,N$.
\end{theorem}

{\noindent\bf Proof:}
The proof is organized into three parts. Part a) proves that the coupled observers in~\eqref{v0-estimation}-\eqref{f0-estimation-2nd-order} yield bounded-error estimation of $v_0$ and $f_0$; Part b)  shows that the observers in~\eqref{vi-estimation}-\eqref{fi-estimation-2nd-order} lead to bounded errors when estimating $v_i$ and $f_i$; finally, based on Parts a) and b), Part c)  demonstrates the upper boundedness of the position and velocity tracking errors when the proposed controller is applied.

{\em Part a):} Define the estimation errors of the observers in~\eqref{v0-estimation}-\eqref{f0-estimation-2nd-order} as $e_{0v,i}=\hat{v}_{0,i}-v_0$ and $e_{0f,i}=\hat{f}_{0,i}-f_0$. According to~\eqref{leader-follower-dynamics-second} and~\eqref{v0-estimation}-\eqref{f0-estimation-2nd-order}, their dynamics can be written as
\begin{align*}
\dot{e}_{0v,i}
=&-b_ie_{0v,i}-\sum_{j \in \mathcal{N}_{i}}a_{ij}(\hat{v}_{0,i}-\hat{v}_{0,j})
+\hat{u}_{0,i}-u_0+\hat{f}_{0,i}-f_0,\\
\dot{e}_{0f,i}
=&-b_ie_{0v,i}-\sum_{j \in \mathcal{N}_{i}}a_{ij}(\hat{v}_{0,i}-\hat{v}_{0,j})-\dot{f}_0.
\end{align*}
Defining  $e_{0vf}=\left[\begin{matrix} e_{0v,1} & \ldots & e_{0v,N} & e_{0f,1}&\ldots & e_{0f,N}\end{matrix}\right]^{\top}$, we have
\begin{align}\label{e0v-dynamics}
\dot{e}_{0vf}=Q_1e_{0vf}+\ell_1,
\end{align}
where
\begin{align*}
Q_1&=\left[\begin{matrix}-H&I\\-H&0\end{matrix}\right],\
\ell_1 =\left[\begin{matrix}e_{0u}\\-\dot{f}_0\mathbf{1}\end{matrix}\right].
\end{align*}
The characteristic polynomial of $Q_1$ is given by
\begin{align*}
\det(sI-Q_1)&=\det\left(\left[\begin{matrix} sI+H&-I\\H&sI  \end{matrix}\right]\right)
=\det(s^2I+ H  s+ H )
=\prod_{i=1}^{N}\left[s^2+\lambda_{i}(H) s+\lambda_{i}(H) \right].
\end{align*}
As is seen from above, the poles of $Q_1$ is stable since $H$ is positive definite.
Then, there must exist a positive definite matrix $P_1$ such that
\begin{align*}
P_1Q_1+Q_1^{\top}P_1=-I.
\end{align*}
For~\eqref{e0v-dynamics},  take a Lyapunov function $V_4(e_{0vf})=\frac{1}{2}e_{0vf}^\top P_1 e_{0vf}$. Consider its derivative:
\begin{align*}
\dot{V_4}&=-\frac{1}{2}e_{0vf}^\top (P_1Q_1+Q_1^{\top}P_1)e_{0vf}+e_{0vf}^\top P_1\ell_1 \\\nonumber
&\leq -\frac{1}{2}\|e_{0vf}\|^{2}+\|e_{0vf}\|\|P_1\|\|\ell_1\|\\
&\leq -\frac{1}{2}\|e_{0vf}\|^{2}+\sqrt{e_u^{2}(0)+q_0^{2}}\|P_1\|\|e_{0vf}\|,
\end{align*}
where  the fact suggested by Lemma~\ref{u0-analysis}  that  $e_u$  exponentially decreases to zero is used. The above inequality can be written equivalently as
\begin{align*}
\dot{V_4}\leq -\alpha_{1} V_4+\beta_{1} \sqrt{V_4},
\end{align*}
with $\alpha_{1}=1/{\bar{\lambda}(P_1)}$ and $\beta_{1}=\sqrt{2(e_u^{2}(0)+q_0^{2})}\|P_1\| / {\underline{\lambda}(P_1)}$.
Hence,
\begin{align}\label{V0vf-inequality}
\sqrt{V_4(t)}&\leq \sqrt{V_4(0)}e^{-\frac{\alpha_{1} t}{2}}+\frac{\beta_{1}}{\alpha_{1}}
(1-e^{-\frac{\alpha_{1} t}{2}})\leq \sqrt{V_4(0)}+\frac{\beta_{1}}{\alpha_{1}}.
\end{align}
It then follows from~\eqref{V0vf-inequality} that
\begin{align*}
&\|e_{0vf}(t)\|\leq \sqrt{\frac{\bar{\lambda}(P_1)}{\underline{\lambda}(P_1)}}\|e_{0vf}(0)\|
+\frac{\beta_{1}}{\alpha_{1}} =\sqrt{\frac{\bar{\lambda}(P_1)}{\underline{\lambda}(P_1)}}\|e_{0vf}(0)\|+
\frac{\bar{\lambda}(P_1)\sqrt{2(e_u^{2}(0)+q_0^{2})}\|P_1\|}{\underline{\lambda}(P_1)},\\
&\lim\limits_{t\rightarrow \infty} \|e_{0vf}(t)\| \leq \frac{\sqrt{2}\bar{\lambda}(P_1)q_0\|P_1\|}{\underline{\lambda}(P_1)}.
\end{align*}

{\em Part b):} Consider the observers for $v_i$ and $f_i$. Define their respective estimation errors as $e_{v,i}=\hat{v}_i-v_i$ and $e_{f,i}=\hat{f}_i-f_i$. Their dynamics can be described as
\begin{align*}\nonumber
\dot{e}_{v,i}
&=-le_{v,i}+e_{f,i},\\ \nonumber
\dot{e}_{f,i}
&=-le_{v,i}-\dot{f}_i.
\end{align*}
Defining $e_{vf}=\left[\begin{matrix} e_{v,1} & \ldots &  e_{v,N} &  e_{f,1} &  \ldots & e_{f,N}\end{matrix}\right]^{\top}$, we have
\begin{align*}\label{ev-dynamics}
\dot{e}_{vf}=Q_2e_{vf}+\ell_2,
\end{align*}
where
\begin{align*}
Q_2&=\left[\begin{matrix}-lI&I\\-lI&0\end{matrix}\right],\
\ell_2 =\left[\begin{matrix}0 & -\dot{f}_1 & -\dot{f}_2 & \cdots & -\dot{f}_N\end{matrix}\right]^\top.
\end{align*}
Following similar lines  to Part a), we  can obtain that   $e_{vf}$ is upper bounded:
\begin{align*}
\|e_{vf}(t)\|&\leq \sqrt{\frac{\bar{\lambda}(P_2)}{\underline{\lambda}(P_2)}}\|e_{vf}(0)\|+\frac{\beta_2}{\alpha_2}
=\sqrt{\frac{\bar{\lambda}(P_2)}{\underline{\lambda}(P_2)}}\|e_{vf}(0)\|+
\frac{\sqrt{2}\bar{\lambda}(P_2)q_1\|P_2\|}{\underline{\lambda}(P_2)},\\
\lim\limits_{t\rightarrow \infty} \|e_{vf}(t)\| & \leq \frac{\sqrt{2}\bar{\lambda}(P_2)q_1\|P_2\|}{\underline{\lambda}(P_2)},
\end{align*}
where $P_2$ is a positive definite matrix  satisfying $P_2Q_2+Q_2^{\top}P_2=-I$.

{\em Part c):} Based on   Parts a) and b), now let us move on to analyze the tracking performance under the controller in~\eqref{controller-second}. Note that the position and velocity tracking errors are governed by
\begin{align*}
\dot{x}_{i}-\dot{x}_{0}=&v_i-v_{0},\\ \nonumber
\dot{v}_{i}-\dot{v}_{0}
=&-k_{1}\left[\sum_{j \in \mathcal{N}_{i}}a_{ij}(x_i-x_j)
+b_{i}(x_i-x_0)\right]
-(v_i-v_{0})
-(\hat{v}_{i}-v_i)
 +(\hat{v}_{0,i}-v_0)\\
&+\hat{f}_{0,i}-f_0
 +f_i-\hat{f}_i+\hat{u}_{0,i}-u_{0},
\end{align*}
for $i=1,2,\ldots,N$.
Define
\begin{align*}
e=\left[\begin{matrix}x_1-x_0 & \cdots & x_N-x_0  &  v_1-v_0 & \cdots & v_N-v_0\end{matrix}\right]^{\top}.
\end{align*} Then,
\begin{align}\label{e-dynamics-second}
\dot{e}=Q_3e+\ell_3,
\end{align}
where
\begin{align*}
Q_3&=\left[\begin{matrix}0&I\\-k_1H&-I\end{matrix}\right],\
\ell_3 =\left[\begin{matrix}0\\-e_v+e_{0v}+e_{0f}-e_{f}+e_u\end{matrix}\right].
\end{align*}
The characteristic polynomial of $Q_3$ is
\begin{align*}
\det(sI-Q_3)&=\det\left(\left[\begin{matrix} sI&-I\\k H&sI+I   \end{matrix}\right]\right)
=\det(s^2I+sI+k H)
=\prod_{i=1}^{n}\left[s^2+s+k \lambda_{i}(H) \right].
\end{align*}
It is seen from above that $Q_3$ is stable because $H$ is positive definite and $k>0$.
If $Q_3$ is stable, there exists a positive definite matrix $P_3$ such that
\begin{align*}
P_3Q_3+Q_3^{\top}P_3=-I.
\end{align*}
Define $V_5(e)=\frac{1}{2}e^\top P_3 e$ for~\eqref{e-dynamics-second}. Then,
\begin{align*}\nonumber
\dot{V_5}&=-\frac{1}{2}e^\top (P_3Q_3+Q_3^{\top}P_3)e+e^\top P_3\ell_3 \\\nonumber
&\leq -\frac{1}{2}\|e\|^{2}+\|e\| \cdot  \|P_3\| \cdot  \|\ell_3\|\\
&\leq -\frac{1}{2}\|e\|^{2}+(\sqrt{2}\|e_{vf}\|+\sqrt{2}\|e_{0vf}\|+\|e_u\|)\cdot \|P_3\| \cdot  \|e\|.
\end{align*}
It can be rewritten as
\begin{align*}
\dot{V_5}\leq -\alpha_{3} V_5+\beta_{3} \sqrt{V_5},
\end{align*}
where $\alpha_{3}=1/{\bar{\lambda}(P_3)}$ and $\beta_{3}=\left(2\|e_{vf}\|+2\|e_{0vf}\|+\sqrt{2}\|e_u\|\right)
\|P_3\|/{\underline{\lambda}(P_3)}$.
Therefore, we have
\begin{align*}
\sqrt{V_5}\leq \sqrt{V_5(0)}e^{-\frac{\alpha_{3} t}{2}}+\frac{\beta_{3}}{\alpha_{3}}
(1-e^{-\frac{\alpha_{3} t}{2}})\leq \sqrt{V_5(0)}+\frac{\beta_{3}}{\alpha_{3}}.
\end{align*}
It then follows that $e$ satisfies
\begin{align}\label{e-boundedness-second-order} \nonumber
&\|e(t)\|\leq \sqrt{\frac{\bar{\lambda}(P_3)}{\underline{\lambda}(P_3)}}\|e(0)\|+\frac{\beta_{3}}{\alpha_{3}}
 \leq \sqrt{\frac{\bar{\lambda}(P_3)}{\underline{\lambda}(P_3)}}\|e(0)\|+
\frac{\bar{\lambda}(P_3)\|P_3\|}{\underline{\lambda}(P_3)}
\left(2\sqrt{\frac{\bar{\lambda}(P_2)}{\underline{\lambda}(P_2)}}\|e_{vf}(0)\|\right.\\
& \quad  \left.+
\frac{2\sqrt{2}\bar{\lambda}(P_2)q_1\|P_2\|}{\underline{\lambda}(P_2)}
+2\sqrt{\frac{\bar{\lambda}(P_1)}{\underline{\lambda}(P_1)}}\|e_{0vf}(0)\|+
\frac{2\bar{\lambda}(P_1)\sqrt{2(e_u^{2}(0)+q_0^{2})}\|P_1\|}{\underline{\lambda}(P_1)}
+\sqrt{2}\|e_u(0)\|\right),  \\ \label{e-limit-second-order}
&\lim\limits_{t\rightarrow \infty} \|e(t)\| \leq \frac{\bar{\lambda}(P_3)\|P_3\|}{\underline{\lambda}(P_3)}\left(
\frac{2\sqrt{2}\bar{\lambda}(P_2)q_1\|P_2\|}{\underline{\lambda}(P_2)}
+\frac{2\sqrt{2}\bar{\lambda}(P_1)q_0\|P_1\|}{\underline{\lambda}(P_1)}\right).
\end{align}
By~\eqref{e-boundedness-second-order}-\eqref{e-limit-second-order}, there exist $\delta$ and $\epsilon$ such that~\eqref{position-velocity-tracking-boundedness}-\eqref{position-velocity-tracking-bounded-limit} hold. This completes the proof.
\hfill$\bullet$

Theorem~\ref{ui-analysis-second} reveals that, for a second-order MAS,  the proposed observer-based controller  can enable a follower to track the leader with bounded position and velocity  errors  when the disturbances are bounded in rates of change. Such an effectiveness is mainly attributed to the proposed observers, through which a follower can estimate the disturbance and velocity variables for tracking control. Further, similar to Remark~\ref{Perfect-error-tracking-first-order}, the tracking error $e_i(t)\rightarrow 0$ as $t\rightarrow \infty$ if  $\dot{f}_i(t) \rightarrow 0$.

\begin{figure*}[t]
\centering
\hspace{13mm}
\subfigure[]{
\raisebox{0.05\textwidth}{%
                  \includegraphics[scale=0.35]{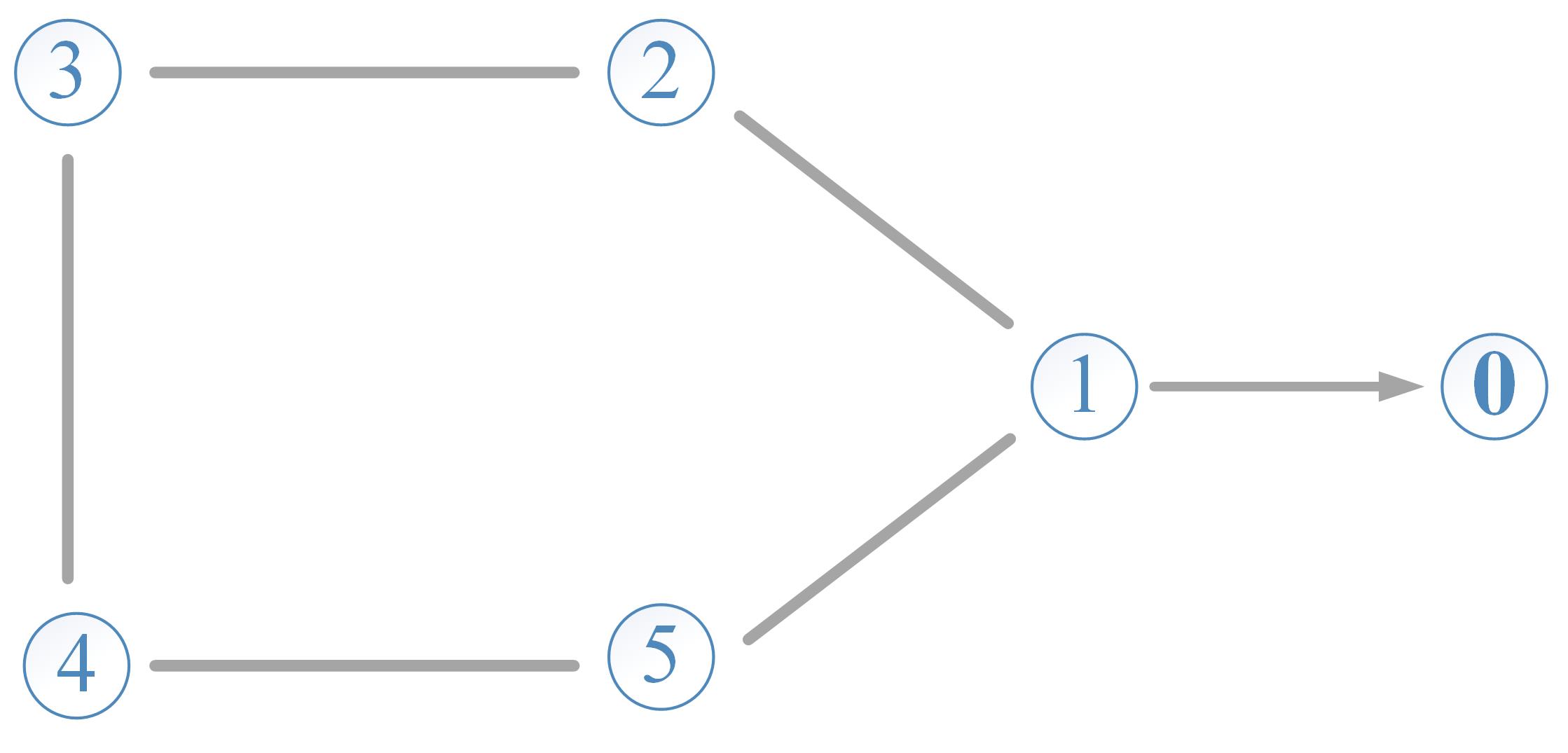}\label{topology}
                }
}\hspace{6mm}
\subfigure[]{
\includegraphics[trim={6mm 16mm 11mm 20mm},clip,width=0.38\linewidth]{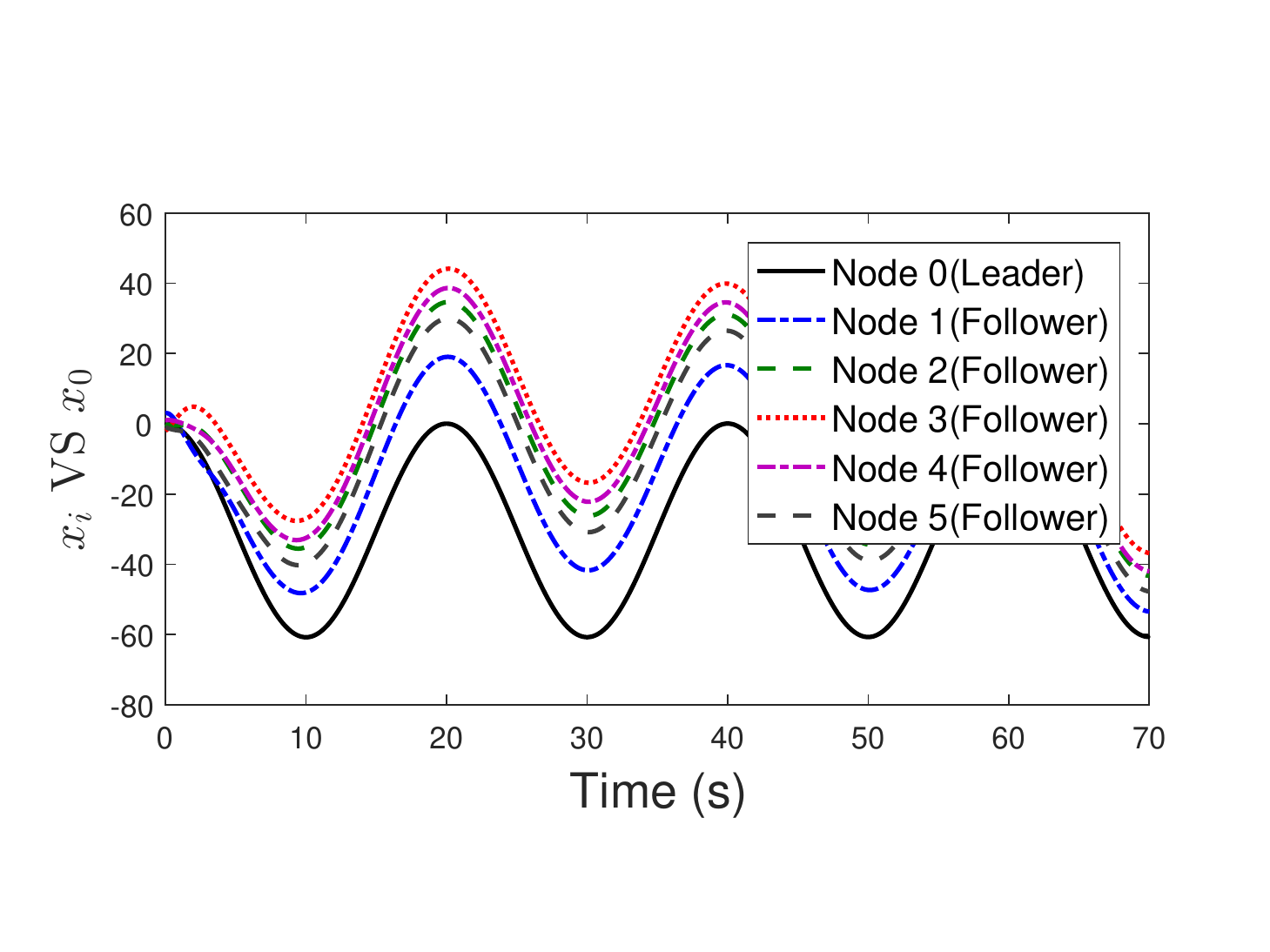}\label{x-second}}\\
\subfigure[]{
\includegraphics[trim={5mm 16mm 12mm
 20mm},clip,width=0.38\linewidth]{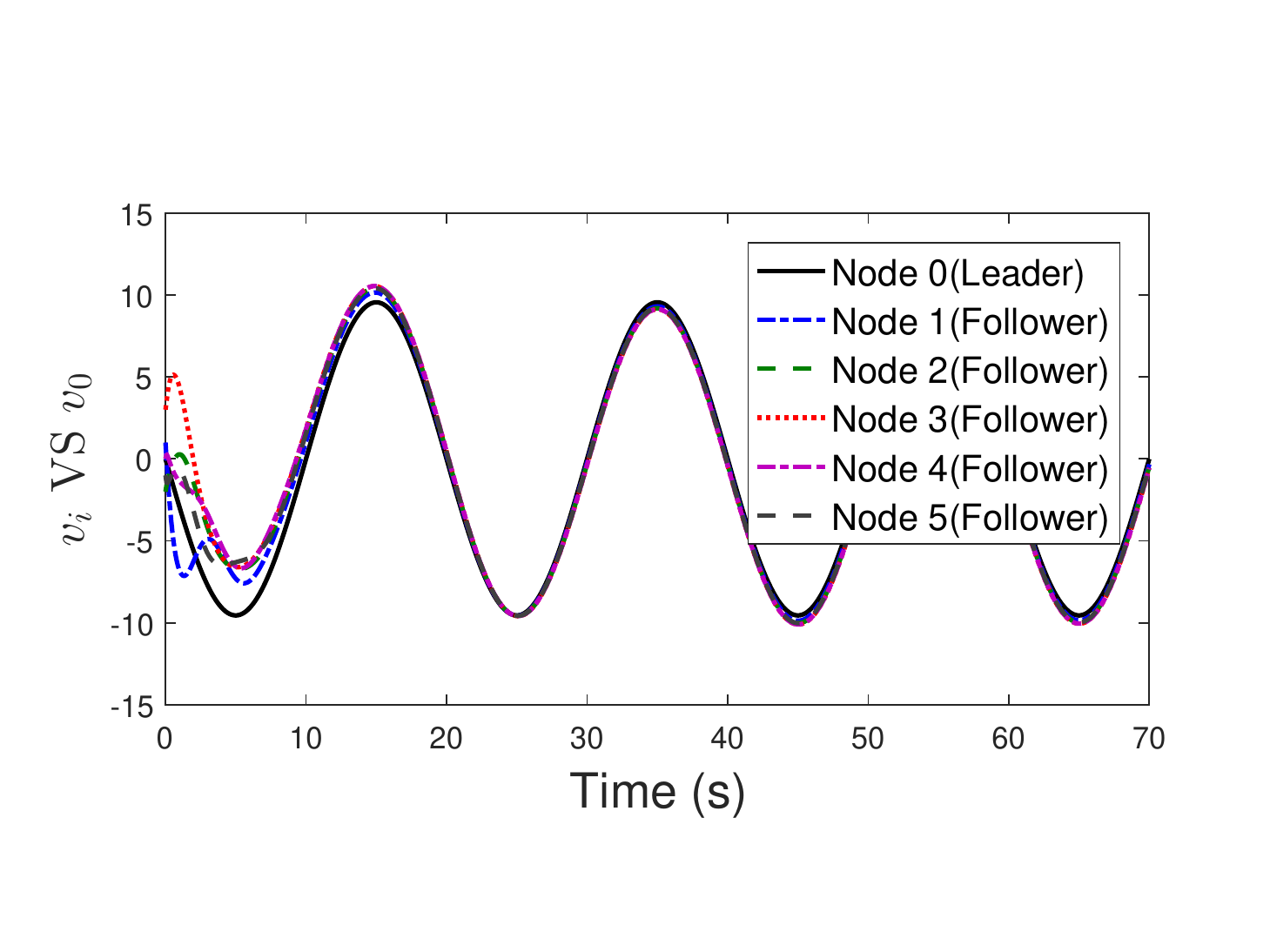}\label{velocity-second}}
 \subfigure[]{
\includegraphics[trim={5mm 16mm 11mm
20mm},clip,width=0.38\linewidth]{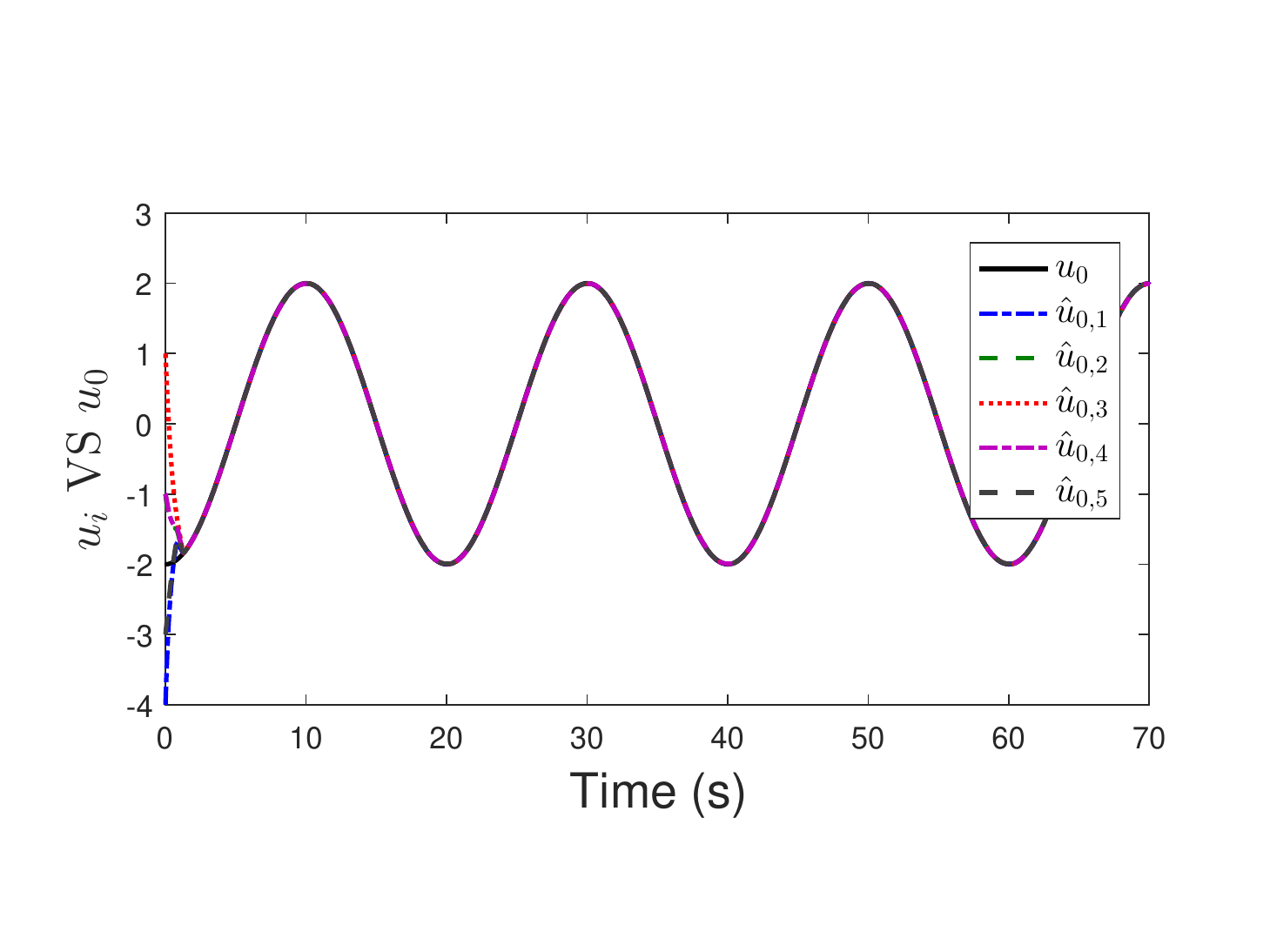}\label{u0-second}}\\
 \subfigure[]{
\includegraphics[trim={6mm 16mm 11mm 20mm},clip,width=0.38\linewidth]{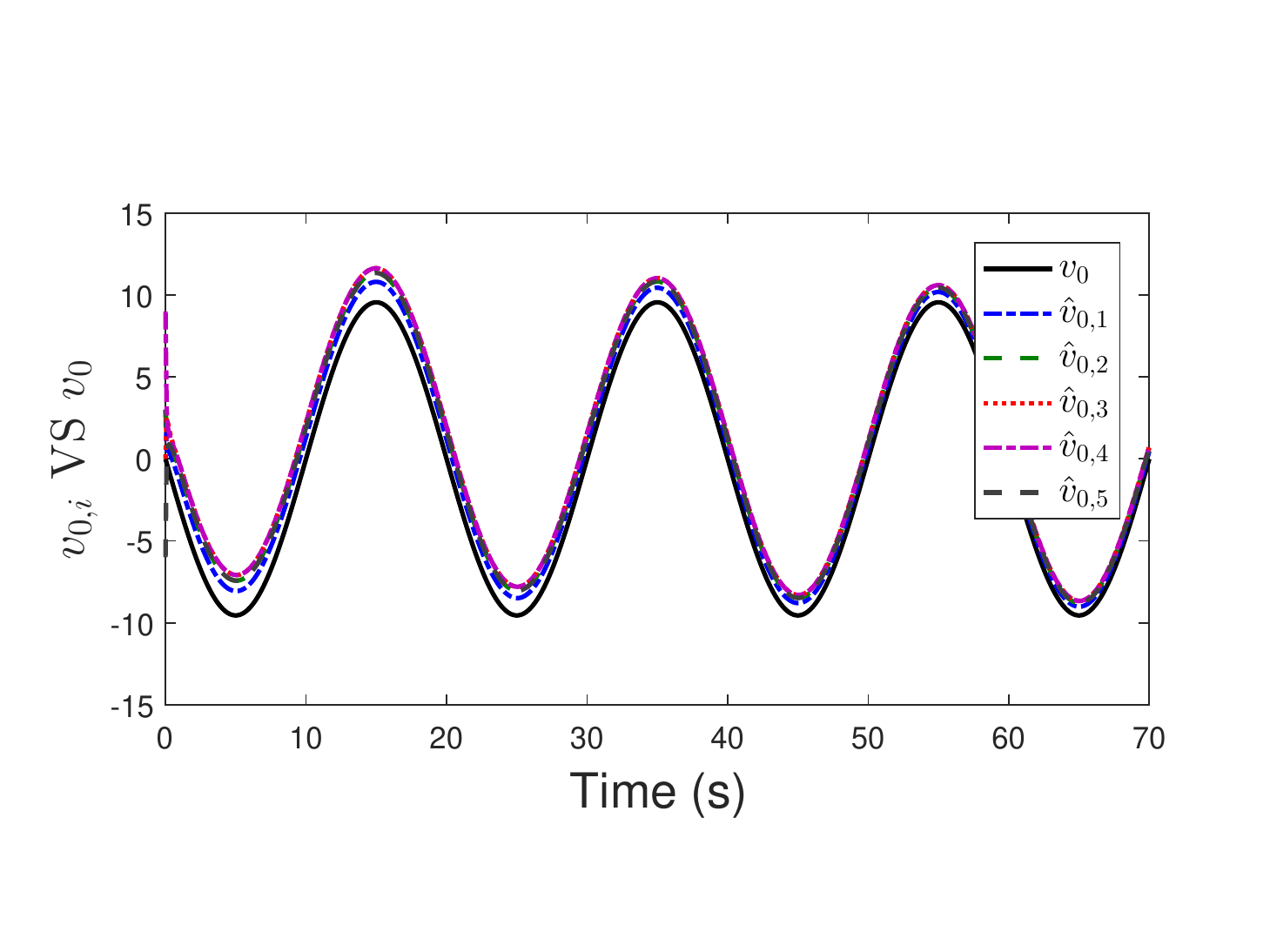}\label{leader-velocity-second}}
 \subfigure[]{
\includegraphics[trim={5mm 16mm 11mm
20mm},clip,width=0.38\linewidth]{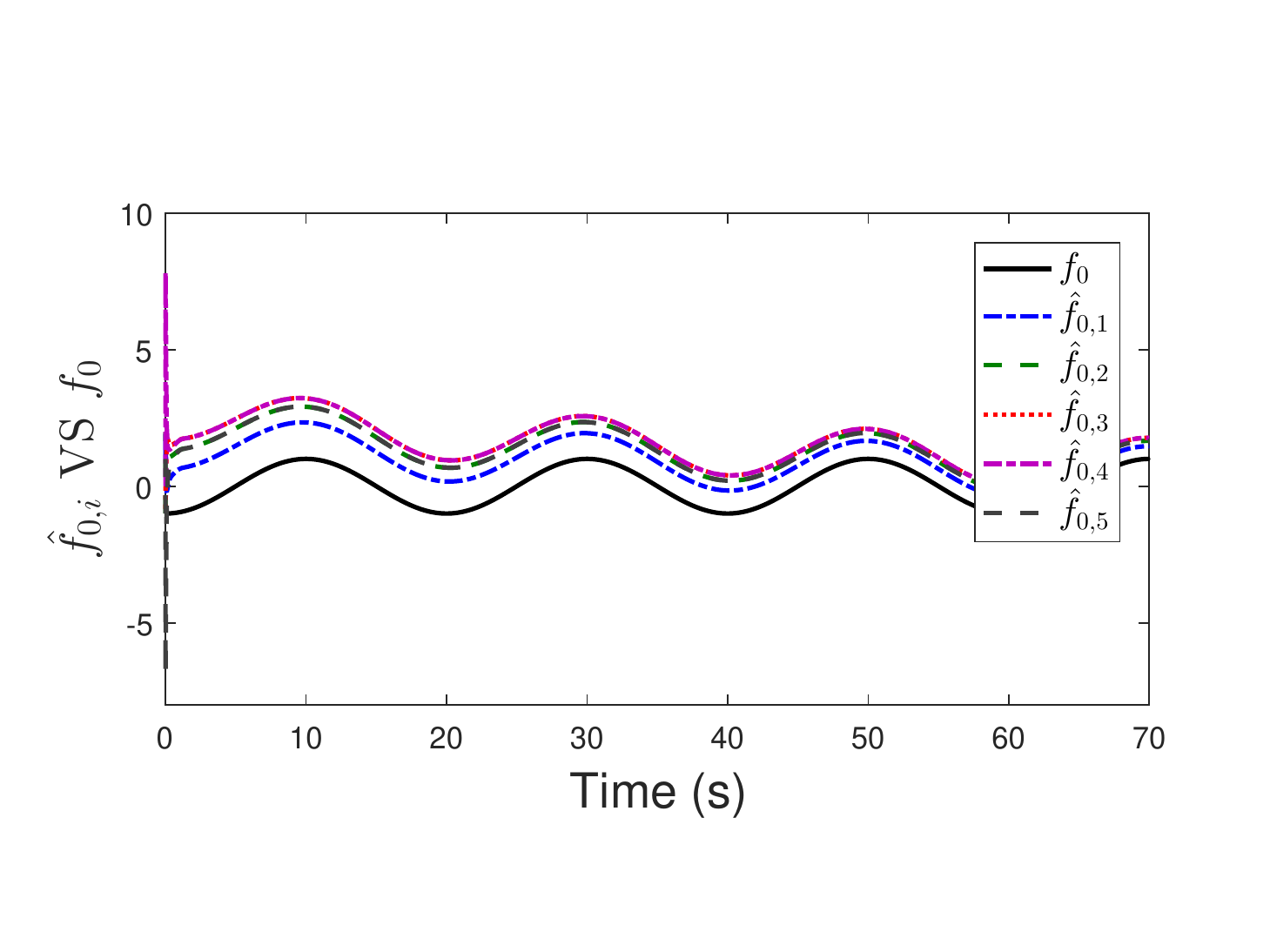}\label{f0-second}}\\
\subfigure[]{
\includegraphics[trim={5mm 16mm 12mm
 20mm},clip,width=0.38\linewidth]{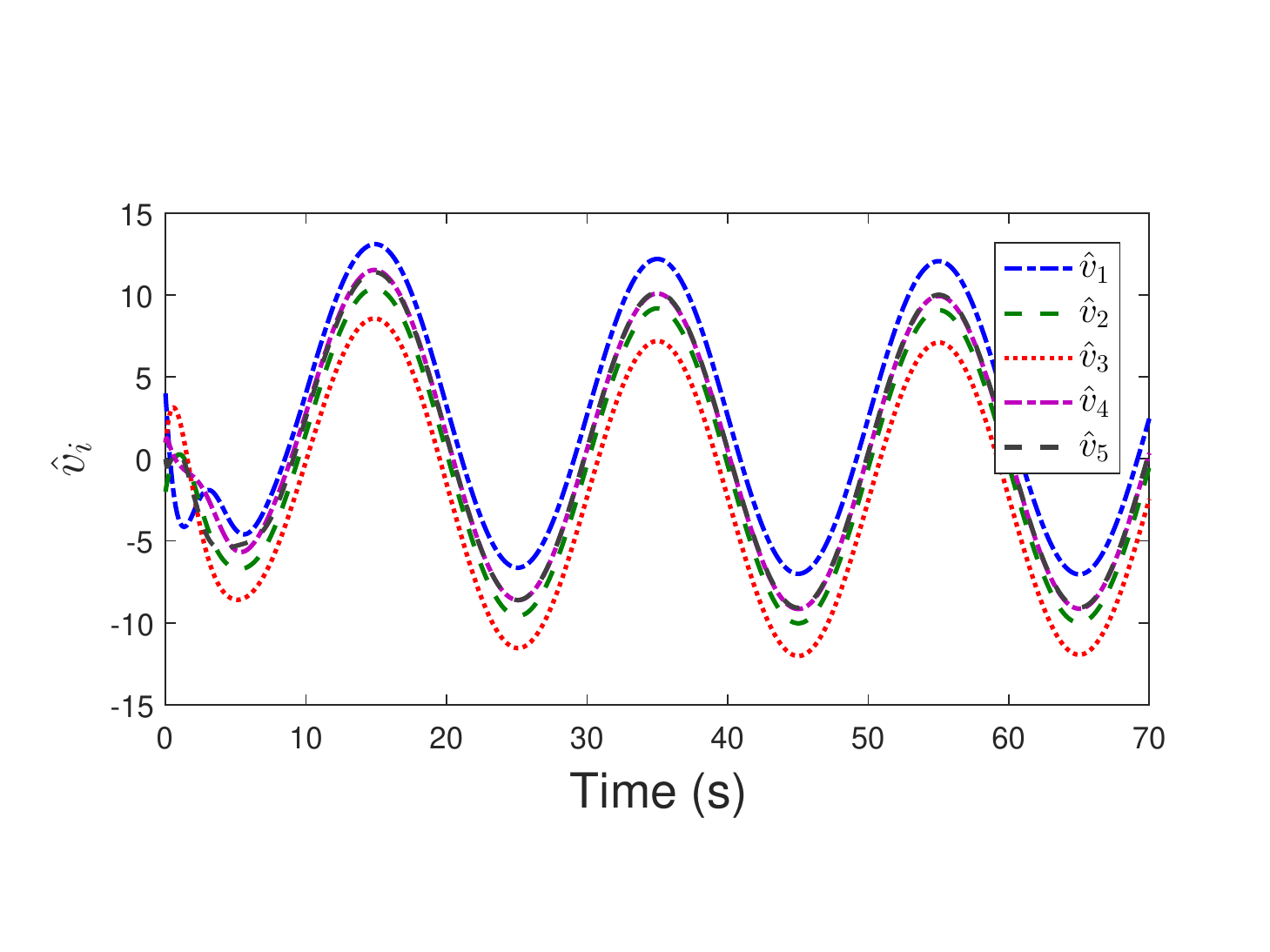}\label{follower-velocity-second}}
\subfigure[]{
\includegraphics[trim={5mm 16mm 12mm
 20mm},clip,width=0.38\linewidth]{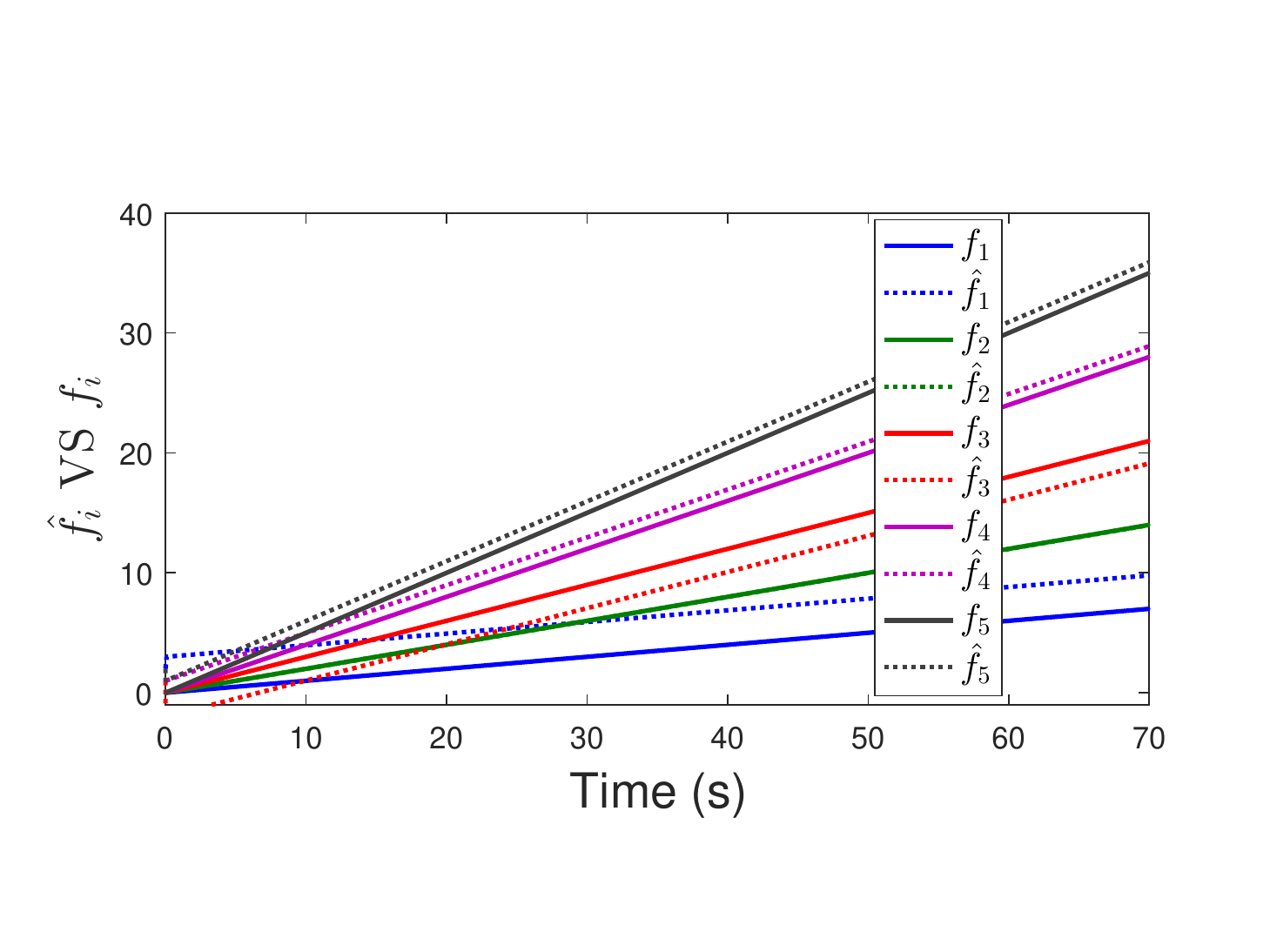}\label{fi-second}}
\vspace {-2mm}
\caption{Second-order MAS tracking control: (a) communication topology; (b) leader's and followers' position trajectory profiles; (c) leader's and followers' velocity profiles; (d) leader's acceleration profile and the estimation by each follower; (e)  leader's velocity profile and the estimation by each follower;
(f) leader's disturbance profile and the estimation by each follower; (g) followers' estimation of their own velocities; (h) followers' disturbance profiles and the estimation on their own.}\label{Second-order-MAS-profiles}
\end{figure*}

\section{NUMERICAL STUDY}\label{simulation}

This section presents an illustrative simulation example     to validate the proposed results. We consider a second-order MAS consisting of one leader and five followers, which share a communication topology shown in Figure~\ref{topology}. Vertex 0 is the leader, and vertices numbered from 1 to 5 are followers. The leader will only send information updates to follower 1, which is its only neighbor. The followers maintain bidirectional communication with their neighbors. For the topology graph, the edge-based weights are set to be unit for simplicity. The corresponding Laplacian matrix  $L$ is then given as follows:
\begin{align*}
L=\left[\begin{matrix}2 &-1& 0 &0 &-1\\
-1& 2& -1& 0& 0\\
0& -1& 2& -1& 0 \\
0& 0& -1& 2& -1\\
-1& 0& 0& -1& 2
\end{matrix}\right].
\end{align*}
Based on the communication topology, the leader adjacency matrix  is
$
B=\mathrm{diag} (1, 0, 0, 0, 0)$. We choose $l=1$ and $k =0.5$, respectively.
The initial conditions include
\begin{align*}
x(0)&=\left[\begin{matrix} 0&3& 0 &-2& 1 &-1   \end{matrix}\right]^\top,\ \
v(0)=\left[\begin{matrix} 0&1 &-2 &3& 0 &-1   \end{matrix}\right]^\top.
\end{align*}
Further,
\begin{align*}\label{u0}
u_0(t)=-2\cos(0.1\pi t),\  \
f_0(t)=-\cos(0.1\pi t),\  \
f(t)=\left[\begin{matrix} 0.1&0.2& 0.3 &0.4& 0.5    \end{matrix}\right]^\top t.
\end{align*}
Note that the disturbances enforced on the followers are bounded in rates of change but linearly diverge through time. This extreme setting is used to illustrate the effectiveness of disturbance rejection here. Apply the observer-based control approach in Section~\ref{second-order} to the MAS.
 The observer-based control approach in Section~\ref{second-order} is applied, with the simulation results outlined in Figure~\ref{Second-order-MAS-profiles}. Figure~\ref{x-second} and~\ref{velocity-second} demonstrate that the followers maintain bounded position and velocity tracking errors, which is in agreement with the results in Theorem~\ref{ui-analysis-second}. It is shown in Figure~\ref{u0-second} that the observer for $u_0$ can gradually achieve accurate estimation through time. This is because the leader can send $u_0$ to its neighbor follower $i$, and with the implicit information propagation, the other followers can eventually  estimate $u_0$ precisely. By comparison, the estimation of $v_0$, $f_0$, $v_i$ and $f_i$ is less accurate, as is seen in Figures~\ref{leader-velocity-second}-\ref{fi-second}, because there is no measurement of them available. However, the differences or estimation errors are still bounded, matching the expectation as suggested by the theoretical analysis.

\section{Conclusion}\label{conclusion}

MASs have attracted significant research interest in the past decade due to their increasing applications. In this paper, we have studied leader-follower tracking for the first- and second-order MASs with unknown disturbances. Departing from the literature, we have considered a much less restrictive setting about disturbances. Specifically, disturbances can be applied to all the leader and followers and assumed to be bounded just in rates of change. This considerably relaxes the usual setting that only followers are affected by magnitude-bounded disturbances. To solve this problem, we have developed  observer-based tracking control approaches, which particularly included the   design of novel  distributed disturbance observers for followers to estimate the leader's unknown disturbance. We have proved that the proposed approaches can enable bounded-error tracking in the considered disturbance setting.   Simulation results further demonstrated the effectiveness of the proposed approaches.

The proposed framework and methodology  provide an ample  scope for  future work. They can be extended to more sophisticated MASs, such as those with nonlinear dynamics, multiple leaders, or a directed communication topology. In addition,  the communication protocols or malicious cyber attacks are of growing importance for the design of future MASs. It will be an interesting question to  leverage the proposed observer-based framework  to address these challenges.


\balance
\bibliographystyle{elsarticle-num}
\bibliography{MAS-Disturbance}

\end{document}